# Fast Fitting of Reflectivity Data of Growing Thin Films Using Neural Networks

Authors

**Alessandro Greco[a], Vladimir Starostin[a], Christos Karapanagiotis[b], Alexander Hinderhofer[a]\*, Alexander Gerlach[a], Linus Pithan[c], Sascha Liehr[d], Frank Schreiber[a]\* and Stefan Kowarik[d]\***

[a]Institut für Angewandte Physik, University of Tübingen, Auf der Morgenstelle 10, Tübingen, 72076, Germany

[b]Institut für Physik, Humboldt Universität zu Berlin, Newtonstr. 15, Berlin, 12489, Germany

[c] ESRF The European Synchrotron, 71, Avenue des Martyrs, Grenoble, 38000, France

[d] Bundesanstalt für Materialforschung und -prüfung (BAM), Unter den Eichen 87, Berlin, 12205, Germany

Correspondence email: alexander.hinderhofer@uni-tuebingen.de; frank.schreiber@uni-tuebingen.de; stefan.kowarik@uni-graz.at

**Funding information**     Bundesministerium für Bildung und Forschung.

**Synopsis**     Artificial neural networks trained with simulated data are shown to correctly and quickly predict film parameters from experimental X-ray reflectivity curves.

**Abstract**     X-ray reflectivity (XRR) is a powerful and popular scattering technique that can give valuable insight into the growth behavior of thin films. In this study, we show how a simple artificial neural network model can be used to predict the thickness, roughness and density of thin films of different organic semiconductors (diindenoperylene, copper(II) phthalocyanine and α-sexithiophene) on silica from their XRR data with millisecond computation time and with minimal user input or a priori knowledge. For a large experimental dataset of 372 XRR curves, we show that a simple fully connected model can already provide good predictions with a mean absolute percentage error of 8-18% when compared to the results obtained by a genetic least mean squares fit using the classical Parratt formalism. Furthermore, current drawbacks and prospects for improvement are discussed.

Fast Fitting of Reflectivity Data of Growing Thin Films Using Neural Networks

## 1. Introduction

X-ray and neutron reflectometry are well-established analytical techniques for thin film metrology. Reflectivity data provides information about the material density via the scattering length density (SLD), as well as the thickness and interface roughness of thin films on an Å-scale. X-ray reflectivity (XRR) is commonly used for crystalline and amorphous films made by sputtering or molecular beam deposition, but also for self-assembled monolayers, biological thin films and even liquid surfaces (Tolan, 1999; Daillant & Gibaud, 2009; Holý et al., 1999; Neville et al., 2006; Wasserman et al., 1989; Braslau et al., 1988). Furthermore, reflectivity measurements can frequently be performed in real-time, which enables *in situ* studies of film growth (Kowarik et al., 2006, 2009; Woll et al., 2011), which inherently is an non-equilibrium process dominated by highly non-trivial statistics and kinetics (Michely & Krug, 2004; Kowarik, 2017). As a result, important dynamic processes, such as nucleation and diffusion, would be missed by post-growth measurements alone, which makes real-time and *in situ* observations indispensable for capturing transient structures.

In the last years, a range of fast XRR techniques have been developed that can acquire XRR curves on timescales as low as 100 ms (Joress et al., 2018; Lippmann et al., 2016; Mocuta et al., 2018), posing challenges to the data handling if on-line monitoring is required. Some of these methods employ energy-dispersive measurements (Kowarik et al., 2007; Metzger et al., 1994; Mukherjee et al., 2002), which are also used in neutron reflectometry (Cubitt et al., 2018). These techniques allow the measurement of large $q$-ranges in one shot while maintaining a fixed scattering angle which increases the data acquisition rate. Moreover, modern high-speed detectors enable the collection of massive amounts data which need to be stored due to the time needed for further treatment and analysis. Clearly, to solve this problem, equally fast analysis tools are desirable that can process data "on-line" and give experiment feedback in real time.

The thickness, roughness and SLD properties of thin films, however, can generally not be extracted directly from reflectivity data, but are instead refined during an iterative fitting process. Various programs are available to accomplish this task by assuming a model for the sample geometry, calculating the resulting Fresnel reflectivity via the Parratt algorithm (Parratt, 1954; Als-Nielsen & McMorrow, 2002) or optical matrix formalism (Heavens, 1955) and iteratively varying the parameters until a good fit is found. Even for a low number of layers, the parameter refinement is laborious and time intensive. Furthermore, a good initial guess of the sample model is often necessary to ensure that the fit converges to a global minimum. Advanced genetic and stochastic fitting algorithms (Björck & Andersson, 2007; Danauskas et al., 2008) are more tolerant towards non-optimal initial parameters and often find a model that fits the measured data, but due to their iterative nature they take much longer than a fast 100 ms XRR curve acquisition. Also, for these algorithms, prior knowledge is needed since there is ambiguity in the interpretation of reflectivity data due to the loss of phase information during the detection process.

Artificial neural networks, or in short neural networks (NN), are an incredibly versatile tool in machine learning that has been applied to a large variety of problems. Their recent widespread use was made possible by the significant increase in computing power by modern graphics cards and specialized neural processing units, as well as the availability of optimized and accessible programming libraries such as TensorFlow (Abadi et al., 2016). NNs already enjoy great popularity in the field of theoretical physics, and their application in physical data analysis has also been successfully demonstrated for a range of methods (Park *et al.*, 2017; Urban III *et al.*, 1998). However, implementations that harness the unique capabilities of machine learning using the performance gain of current programming libraries and graphics cards for experimentalists are so far largely absent.

The goal of this work is to show as a proof of concept that NNs can be used to not only reduce the user input and computation time needed to extract thin film properties from XRR data, but also promise to alleviate the requirement of a priori knowledge about the studied system. This makes NNs ideal for the application in real-time measurements. In this study, we demonstrate the performance of a fully connected NN with six hidden layers trained with simulated XRR data and tested on five real-time XRR datasets of growing organic thin films. However, we emphasize that in principle any material combination is possible. We also discuss possible extensions and limitations of our approach.

## 2. Neural network design

### 2.1. Architectures and training

In this study, we employ a feed-forward neural network using supervised learning with simulated training and validation data. In this architecture, information is processed from a set of input neurons to a set of output neurons through multiple so-called hidden layers of neurons. These sets of neurons are called layers and should not be confused with the same term that is often used to describe the layered structure of thin films. The input layer of the NN represents the measured X-ray intensity values for at different momentum transfer values $q_z$ and the output layer corresponds to the different thin film properties, i.e. film and oxide thickness, roughness and density. A schematic of the architecture used in this study is shown in Figure 1. In the case of fully connected models, such as the one described herein, the value of each neuron after the input layer is calculated by a weighted sum of all neurons in the previous layer. Before passing it on to the next layer, the summed values are passed through an activation function. In our case, we use a linear rectifier function (ReLU) which is a common default setting that performs well on many tasks.

This way, for any given reflectivity curve a corresponding output can be calculated. During the training process, all weights in the network are adjusted so that for an arbitrary set of input values in the training data the correct set of output values are obtained. This is achieved by randomly choosing a small subset of the training data (called minibatch) and calculating the average error between the obtained output and the expected output known from the simulation using a cost function, here the mean squared error. Once the error is determined, a backpropagation algorithm based on stochastic gradient descent is used to determine how the weights in the network must be updated in order to minimize the error (Bottou, 1991; Hecht-Nielsen, 1992). This process is repeated for several full passes through the entire training data, called epochs. The optimization algorithm employed in this work is ADAM (Adaptive Moment Estimation) (Kingma & Ba, 2014).

The neural network model employed in this study (Figure 1) consisted of 6 fully connected hidden layers with 400, 800, 400, 300, 200 and 100 neurons, respectively. For the results discussed in this work, the predictions of three independently trained neural networks with the same hyperparameters and training data, but random initialization were averaged.

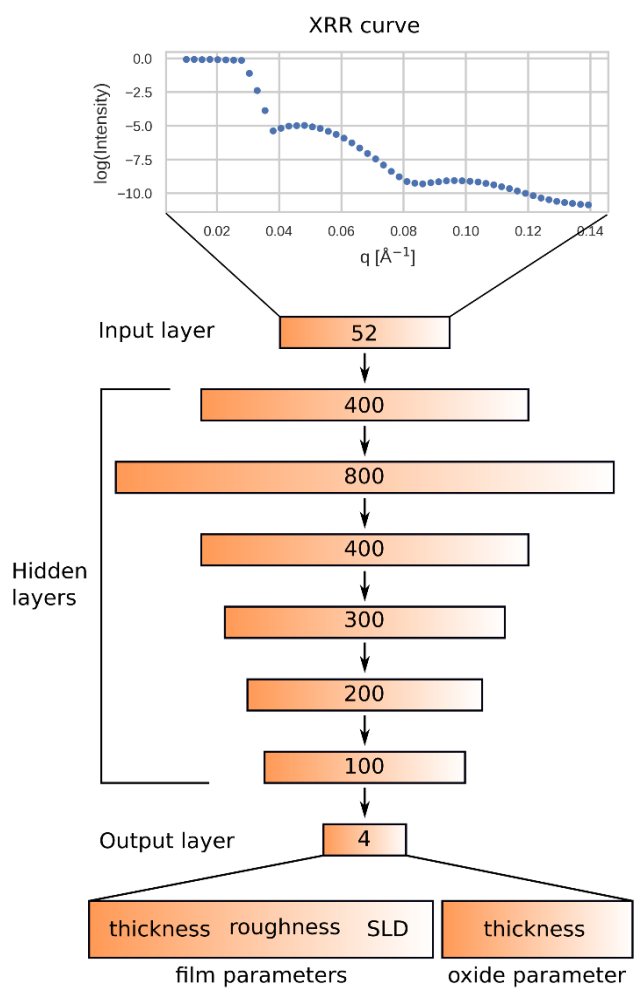

**Figure 1** Schematic of the neural network architecture used in this work. The input layer consists of 52 reflectivity values at discrete $q_z$ positions. The output layer consists of 4 sample parameters: 3 film parameters (thickness, roughness and scattering length density (SLD)) and one substrate parameter (thickness of the native silicon oxide). All layers are fully connected with the next by weights that are randomly initialized and then optimized.

Both the simulated and the experimental data were normalized to 1 and passed through a log function before using them as input. This was done in order to reduce the number of orders of magnitude over which the input data is distributed. A large distribution of input values is a common problem that can inhibit training, since it produces strongly varying weighted sums in the neural network. In more sophisticated approaches, one may consider other weighting or normalization methods. Furthermore, each output parameter of the model was normalized to the minimum and maximum values of the training data so that the mean square error cost function optimizes for all thin film parameters. To keep track of the performance of the model during training and to judge its ability to generalize and make predictions on data that are not included in the training data, its accuracy was evaluated with independently generated validation data. After every epoch, the trained model predicts the outputs of the validation data and the validation error is calculated using the same error function as for the training set. In general, a validation error that is much higher than the training error signifies that the network is overfitting to the training data. On the other hand, if the validation and training error are very similar, the capacity of the model might be too low to capture important features in the data.

The training and validation error shown in Figure 2 are representative of a typical training session of the NN described above. Even though the training and validation loss could be further reduced by an order of magnitude through longer training, we observed lower prediction accuracies on experimental data when the model was trained for more than 60 epochs. The reason for this is that even though we do not see any overfitting with respect to the validation data, there is likely overfitting with respect to the experimental data when the model is trained for too long. Thus, we used the model with the lowest validation loss within 60 epochs to achieve a trade-off between training loss and the ability to generalize to experimental data. While overfitting is a general issue of many machine learning problems, the number of epochs after which it occurs might vary strongly for different types of data and NN architectures. Thus, the optimal number of epochs has to be determined empirically for a given problem and likely depends also on the amount quality of training data and its similarity to the experimental data.

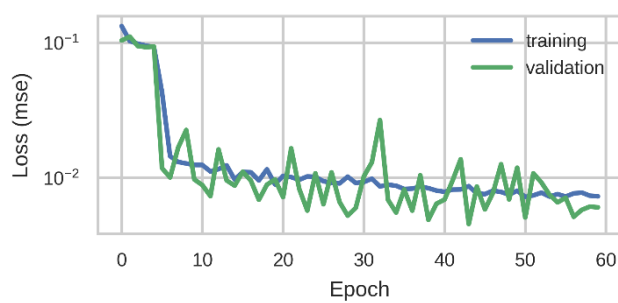

**Figure 2** Characteristic training and validation error during training of the neural network demonstrated in this study. Since the validation error is very close to the training error, there is likely no overfitting with respect to the validation data.

## 2.2. Data preparation

One of the most important factors for the performance of a given neural network architecture is the quality and choice of the training data. It is crucial to have a sufficiently large and varied data set to allow the network to generalize over the entire parameter space. Optimally, a large training dataset of measured data with precisely labelled thin film parameters should be available for training, validation and testing of the NN model. However, since it is unfeasible to perform the necessary amount of independent experiments and fit them manually for classification, we used simulated training and validation data. We simulated 200,000 XRR curves with a 4:1 training/validation split using an adaptation of the optical matrix method (Heavens, 1955; Abelès, 1950), which is a computationally more efficient alternative to the recursive Parratt formalism (Parratt, 1954). For this purpose, parts of the Refl1D source code (Copyright (c) 2006-2011, University of Maryland All rights reserved) were used. Furthermore, we assumed a thin film sample structure with three thin film layers: two for the substrate (silicon and native oxide) plus the deposited thin film. The model for the interface roughness was assumed to have a root mean squares distribution (Névot & Croce, 1980). The roughness of $Si/SiO_x$ substrates is known to be very low and thus, we assumed a constant roughness for the $SiO_x$ and Si layers of 1 and 2.5 Å, respectively. Furthermore, the SLDs of those layers were assumed to be constant with values of 17.8 and $20.1 \times 10^{-6}$ Å$^{-2}$, respectively. The parameters of thickness, SLD, and roughness were uniformly distributed within the generated training data. For the deposited film, the ranges of the thickness and SLD were 20-300 Å and $1-14 \times 10^{-6}$ Å$^{-2}$, respectively. Training data with a thickness below 20 Å was excluded since, due to their ambiguity, they were the most difficult to predict for the NN and by removing them, the prediction accuracy on the rest of the data could be improved. The range of the roughness was up to half the film thickness, but limited to 60 Å. The thickness of the native oxide layer was assumed to be within 3-30 Å. The reflectivity curves were simulated in a $q$ range between 0.01 and 0.14 Å$^{-1}$ at 52 equally spaced points, which is comparable to the resolution of our experimental data. The small $q$-range was chosen to avoid conflicts with Bragg reflections and corresponding Laue oscillations, which are not part of our simple box model.

For performance evaluation of the NN, we used experimentally measured XRR curves of real-time growth of diindenoperylene (DIP), copper(II) phthalocyanine (CuPc) and α-sexithiophene (6T) on silicon substrates with a native oxide layer. Appropriate footprint corrections and normalization was applied to the data before further use. The predictions of the model were judged against a conventional least mean squares (LMS) fit that was performed manually on 20% of the curves. The SLD profiles of each film at their final thickness are shown in section S2 of the supporting information. The rest of the film parameters were linearly interpolated within one measurement. The fit was performed with six open parameters: the thickness, roughness and SLD of the deposited film, the thickness and roughness of the oxide layer and the roughness of the silicon substrate. For CuPc and 6T, we also included a thin void layer with a thickness of 3 Å and a roughness of 1 Å between the substrate and the film. This was done because for some organic thin films, the electron density (and thus SLD) at the interface with the substrate is lower than in the bulk and including a void layer with a finite roughness improves the fit quality. In these cases, the NN model is intentionally simpler than the manual fit, but since the void layer is thin compared to the deposited film, we can directly compare the film thicknesses obtained from both the NN and the LMS fit. The densities of the silicon and its oxide layer were assumed to be constant among all experiments as described above. In order to make all XRR curves compatible with the same fixed size of the input layer, the reflectivity curves of all experiments were interpolated to the same 52 $q$ values without significant change in curve shape.

## 3. XRR fitting performance

To evaluate the prediction accuracy of our NN model, we tested its performance on 20,000 independently simulated curves with the same parameter range as the training data, as well as on each of the five experimental real-time XRR datasets. In the case of the simulated data, the mean average percentage errors of the film thickness, roughness and SLD were 8%, 16% and 6%, respectively. While already quite good, these metrics reveal that for this NN model there is still a significant portion of misclassified curves. Furthermore, within the given $q$-range, correctly determining the roughness seems to be intrinsically more difficult than the other two parameters. Since the synthetic test data was generated using the same process as the training data, we cannot expect better performance on data that was generated using a different process, i.e. experimental data. While a reduction of the training and validation loss could be achieved in principle (for example through longer training sessions), we observed that this generally leads to a decrease of the performance on experimental data. This means, that the training loss alone cannot be used to estimate how the neural network will perform on experimental data and the training process is ultimately limited by the fact that the simulated data does not perfectly match the experimental data.

For the performance evaluation on experimental data, the film properties predicted by the model were compared to a manual LMS fit using a genetic algorithm (GenX). The studied systems were two DIP films, one CuPc film and one 6T film each grown at 303 K as well a third DIP film grown at 403 K. Three out of five of these datasets have already been analyzed and published before (DIP 303K (Hinderhofer *et al.*, 2010), DIP 403K (Kowarik *et al.*, 2006), 6T (Lorch *et al.*, 2015)). Panels a-c of Figure 3 show this comparison for a DIP film grown at 303 K (all other datasets are shown in the supporting information). It is immediately apparent that for most of the series, the predicted values are close to the ones obtained via the manual LMS fit. We note that this achievement is already remarkable, since the network has no concept of any temporal correlation between the XRR curves, which is the kind of knowledge a researcher would use when selecting bounds and starting points for an LMS fit. Furthermore, the parameter predictions of each XRR curve were obtained on average within 77 ms when predicting a single curve and 0.03 ms when predicting 20,000 curves at once.

Fast Fitting of Reflectivity Data of Growing Thin Films Using Neural Networks

Compared to a manual fit, this is orders of magnitude faster and can compete with the speed at which modern 2D detectors operate. Also, after training, there was no additional input necessary. This makes it possible to predict film properties during measurements in real-time without the need of human supervision.

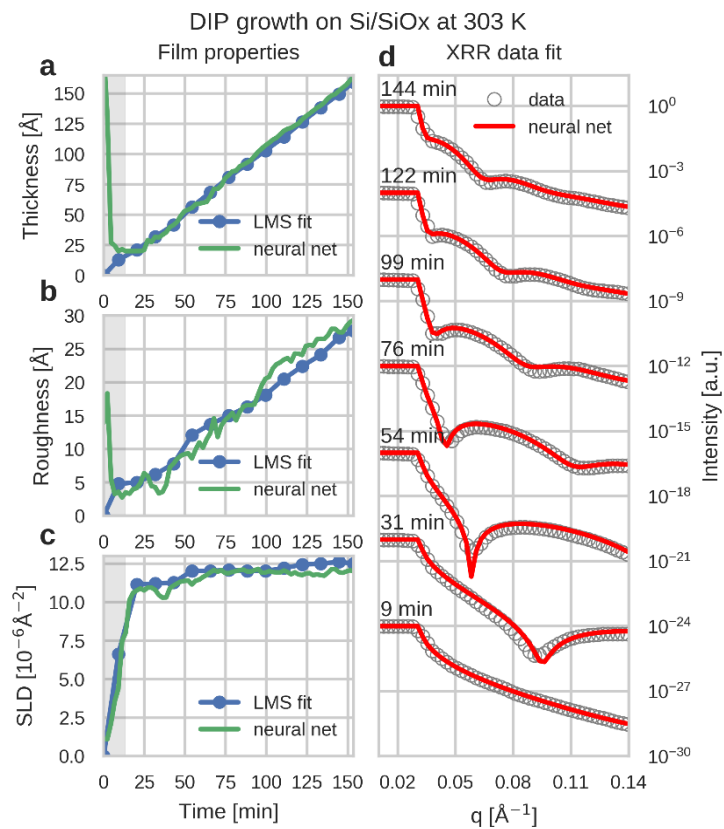

**Figure 3** Fitting performance of the neural network model on a DIP film grown at 303 K with a deposition rate of 1 Å/min. (a-c) Comparison of the film parameters predicted by the neural network with results from least mean square fitting with human supervision at different times during growth. The shaded area marks films with thicknesses below 20 Å, where the network has not been trained and consistently predicts thick films with high roughness. (d) Overlay of the experimental XRR data with data simulated using the parameters predicted by the NN during different times during growth.

Panel d shows an overlay of experimental reflectivity data with simulated curves using the predicted film parameters at different times during growth. In general, the curves show a good s available to neural networks, for example by tuning the range and distribution of the training data. While it is ultimately desirable to also reliably fit these curves using our NN approach, it is clear that any predicton based on data with a higher amount of ambiguity will also have a higher level of uncertainty.

Table 1 shows the mean average percentage error of the NN predictions when compared to the values determined via the manual fit, excluding films with thicknesses below 20 Å. Similar to the results for simulated data, the error is highest for the film roughness and lowest for the SLD. However, on average, the prediction accuracy on experimental data is 2-3 percentage points lower in all three categories. This is likely due to several reasons: Firstly, the accuracy on real data is expected to be lower than on the simulated data since there is already an error attached to the parameters which were extracted via the manual fit before comparison with the NN results. Therefore, the errors of both the LMS and NN fit contribute to the deviation, which is not the case for the simulated test data, where we have perfect knowledge about underlying simulation parameters. Secondly and most importantly, it is probable that the simulated training data differs from the experimentally measured data with regards to a finite experimental resolution and noise and, as a result, the model is trained to subtle features in the simulated data that may be different or not present in the real data. Also, the simulation model with a single film layer may not describe the real system accurately enough or there may be some systematic artefacts of the measurement setup that are difficult to account for in the simulation.

**Table 1** Mean absolute percentage error and standard deviation of the predictions on experimental XRR curves with respect to the values obtained via a conventional LMS fit with manually set bounds and starting points. Predictions of films with a thickness below the training range of the NN (<20 Å) and high roughness (>30 Å) were excluded. DIP 303 K (1) is shown in Figure 3, all others are shown in the supporting information.

|  | DIP 403 K | DIP 303 K (1) | DIP 303 K (2) | CuPc 303 K | 6T 303 K | total |
| --- | --- | --- | --- | --- | --- | --- |
| thickness | 17±20 % | 4±4 % | 6±9 % | 16±13 % | 14±3 % | 11±10 % |
| roughness | 20±14 % | 12±11 % | 15±11 % | 26±18 % | 16±11 % | 18±13 % |
| SLD | 11±9 % | 3±2 % | 9±8 % | 6±5 % | 10±6 % | 8±6 % |

Apart from relying on these metrics, we also confirmed the physical validity of the parameter predictions by taking experimental conditions into consideration. Out of the three parameters, the prediction of the thickness can be verified the easiest, since in all experiments the films were grown at a constant rate. This expected linear behavior is obtained for all experiments and coincides perfectly with the LMS fit. The obtained thickness values can also be verified to a high degree of certainty by considering the periodicity of the Kiessig fringes. The predicted SLD also shows the qualitatively expected behavior of a continuous increase during the beginning of the thin film growth with a saturation at a value that is somewhat lower than the SLD of the solid-state crystal. This indicates the transition from a bare substrate to an organic thin film with a constant in-plane-averaged electron density.

Among the three predicted properties for each experiment, the roughness evolution is arguably the most difficult to judge since it strongly depends on the specific molecular system and on several important experimental parameters, such as the growth rate and the substrate temperature (Michely & Krug, 2004; Kowarik, 2017). In the studied systems, however, we generally expect an overall increase in roughness for higher film thicknesses and this behavior is predicted by the NN model for all shown datasets.

**Conclusions**

In this work, we demonstrated how a straightforward neural network model with fully connected layers can be used to extract the film thickness, roughness and density parameters from real-time reflectivity data of thin films. The small but deep neural network model was trained on simulated data and tested on simulated and experimental data. While the prediction accuracy was lower on experimental data, it still achieved high accuracies with a mean absolute percentage error between 8-18% with respect to the result determined via a manual fit. Importantly, among the three parameters, the film roughness was the most difficult to predict for the model in both the synthetic as well as the experimental data. While the prediction accuracy on synthetic data could in theory be increased by training the model for longer, this so far did not translate to improved accuracy for the experimental predictions. Thus, future efforts should focus on generating better training data that more accurately represent the experiment in order to allow longer training times without overfitting to features that are only present in the simulation. Another possibility is to define a loss function that places a higher weight on the relevant parameters such as the roughness or the parameter range that is prone to high prediction errors. Nevertheless, it is important to understand and improve the prediction behavior on simulated data, since it essentially represents the upper limit of what can be expected in terms of prediction accuracy. Possible strategies for improvement should involve optimizing the quantity, quality and distribution of training data as well as testing more sophisticated neural network models, such as convolutional models, that may more easily capture the required features in the training data. Furthermore, exhaustive optimization of hyperparameters, such as the learning rate or the model capacity, are likely necessary to achieve higher prediction accuracies even for thin or very rough films.

While we expect significant improvements of NN-based models in the future, the current performance is already useful for a preliminary screening of reflectivity data before further analysis. It may also potentially be used directly within the live view of the diffractometer control software. Together with the extremely fast prediction times of 0.03 – 77 milliseconds per curve and the fact that, after training, no further user input is needed, this approach is perfectly suited for *in situ* applications, such as monitoring film parameters during real-time measurements.

We also emphasize that the extension of this model to more complex systems, such as multilayers and layers with internal structures, is in principle possible if appropriate training data is available in sufficient quantity. In the same way, this approach is in principle easily transferrable to neutron reflectivity data, however, some important differences such as the different cross sections of neutrons (coherent as well as incoherent) need to be taken into account. Addressing these complications might require some adjustments to the current neural network architecture, which should be part of future studies.

**Acknowledgements**    We gratefully acknowledge the assistance of Philip Willmott and Oleg Konovalov at the Surface Diffraction beamline at the Swiss Light Source and the ID10 beamline at the European Synchrotron Radiation Facility, respectively. We also acknowledge the help of Takuya Hokosai during data acquisition. Furthermore, we thank Oliver H. Seeck (DESY) for his support. Lastly, we thank Phillip Berens and Matthias Bethge from the Cluster of Excellence in Tübingen for machine learning in science for our fruitful discussions.

Fast Fitting of Reflectivity Data of Growing Thin Films Using Neural Networks

# Supporting information

### S1. Additional results of predictions on experimental data

Figure S1 to Figure S4 demonstrate the prediction performance of the neural network model on the datasets that were used to calculate the predictions accuracy values, but not shown in the main text. In general, the network performs worse on XRR curves that have less pronounced features, such as low thickness or high roughnesses. These curves are marked with shaded areas and were not included in the accuracy statistics.

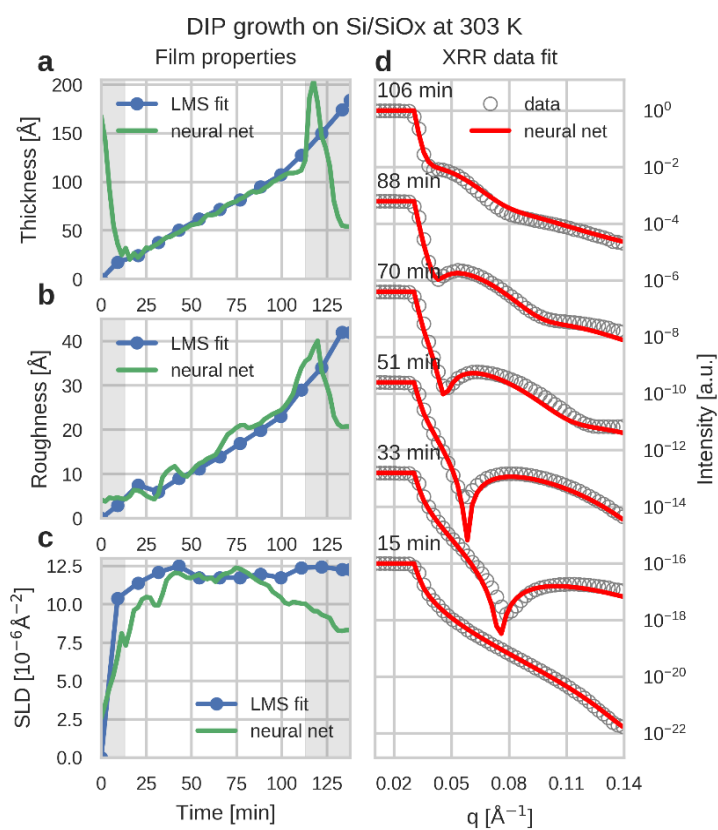

**Figure S1**  Fitting performance of the neural network model on a DIP film grown at 303 K with a deposition rate of 1.3 Å/min. (a-c) Comparison of the film parameters predicted by the neural network with results from least mean square fitting with human supervision at different times during growth. The shaded area marks films with low thickness (below 20 Å) or high roughness (above 30 Å) where data is difficult to fit for the NN. (d) Overlay of the experimental data with data simulated using the parameters predicted by the NN during different times during growth.

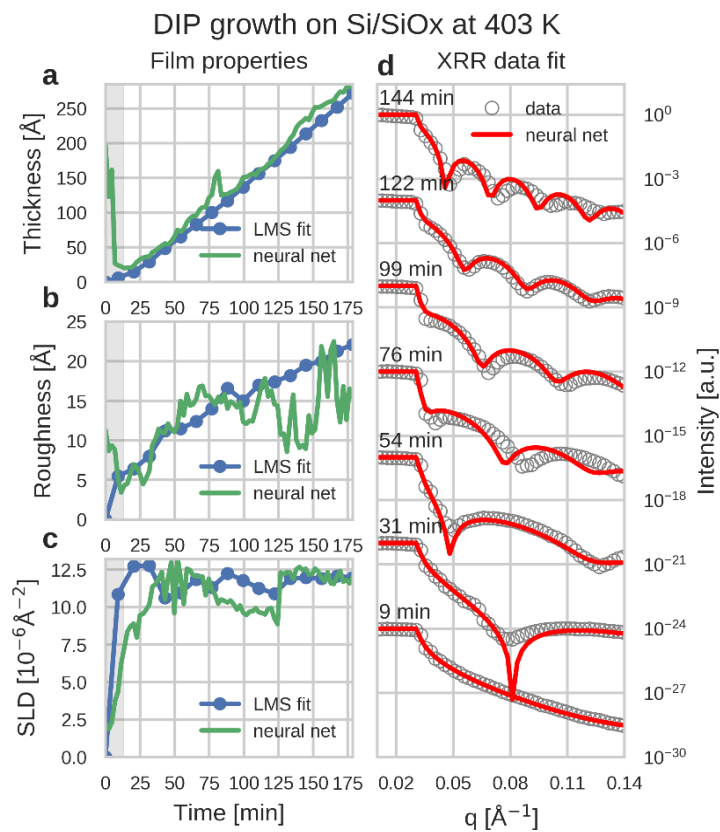

**Figure S2** Fitting performance of the neural network model on a DIP film grown at 403 K with a deposition rate of 1 Å/min. (a-c) Comparison of the film parameters predicted by the neural network with results from least mean square fitting with human supervision at different times during growth. The shaded area marks films with low thickness (below 20 Å) where the data is difficult to fit for the NN. (d) Overlay of the experimental data with data simulated using the parameters predicted by the NN during different times during growth.

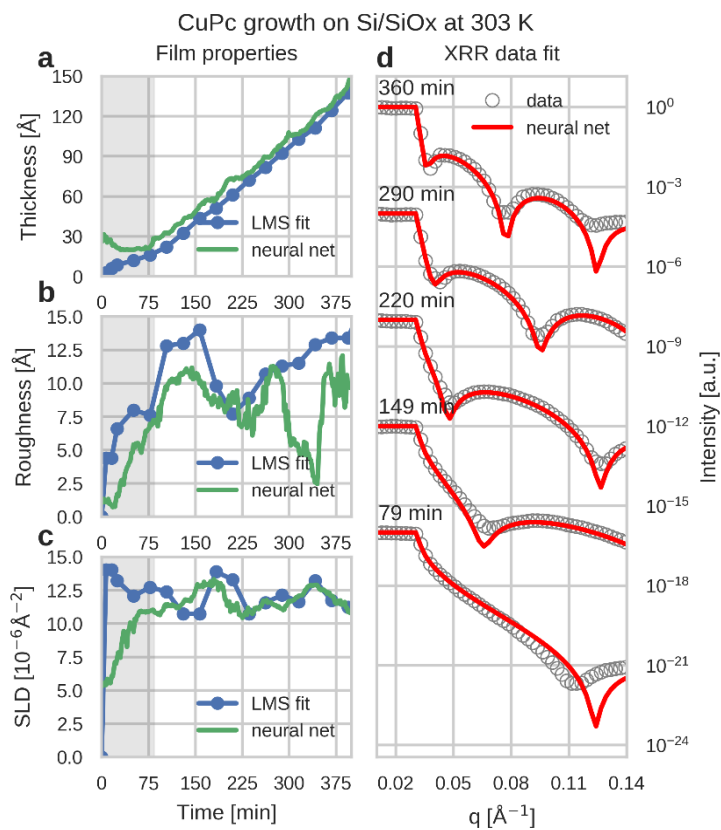

**Figure S3** Fitting performance of the neural network model on a CuPc film grown at 303 K with a deposition rate of 0.4 Å/min. (a-c) Comparison of the film parameters predicted by the neural network with results from least mean square fitting with human supervision at different times during growth. The shaded area marks films with low thickness (below 20 Å) where the data is difficult to fit for the NN. (d) Overlay of the experimental data with data simulated using the parameters predicted by the NN during different times during growth.

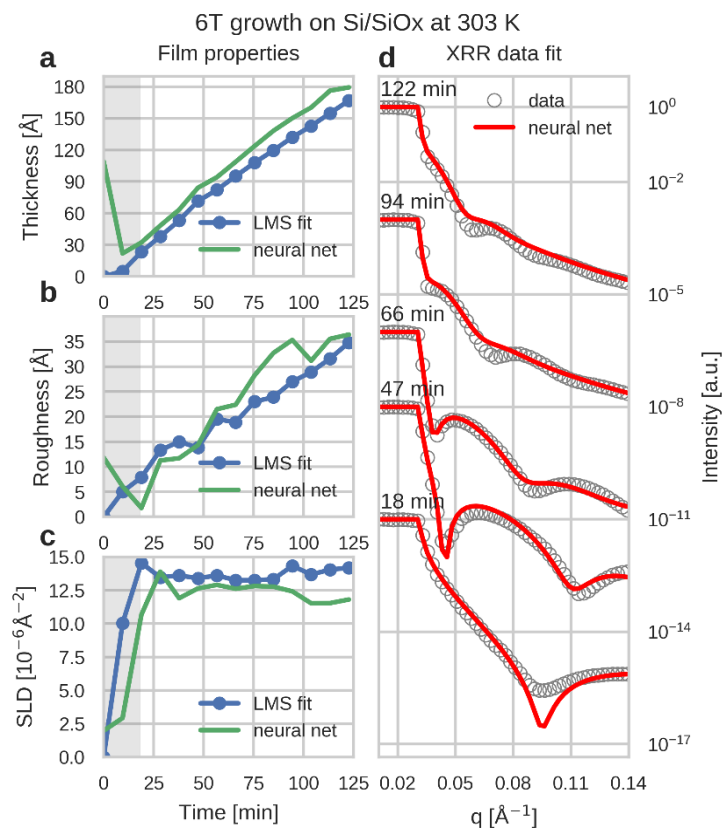

**Figure S4** Fitting performance of the neural network model on a 6T film grown at 303 K with a deposition rate of 1.3 Å/min. (a-c) Comparison of the film parameters predicted by the neural network with results from least mean square fitting with human supervision at different times during growth. The shaded area marks films with low thickness (below 20 Å) where the data is difficult to fit for the NN. (d) Overlay of the experimental data with data simulated using the parameters predicted by the NN during different times during growth.

## S2. Scattering length density profiles for the studied thin films at their final thickness

Figure S5 to Figure S9 show the SLD profiles of the five studied organic thin films at their final thickness, obtained via conventional (manual) least mean squares fitting.

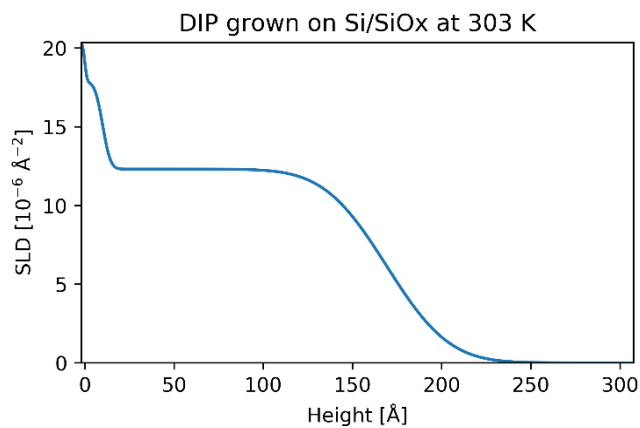

**Figure S5** SLD profile of the DIP film grown on Si/SiO$_x$ at 303 K with a deposition rate of 1 Å/min and a final thickness of about 180 Å.

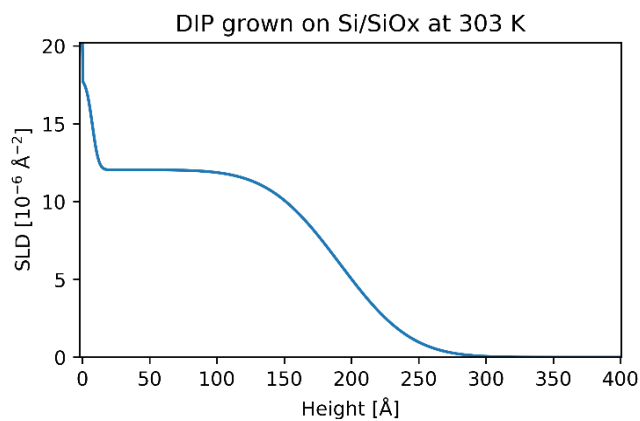

**Figure S6** SLD profile of the DIP film grown on Si/SiO$_x$ at 303 K with a deposition rate of 1.3 Å/min and a final thickness of about 170 Å.

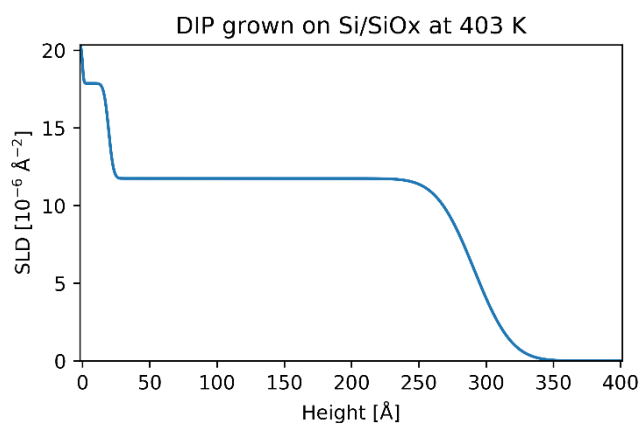

**Figure S7** SLD profile of the DIP film grown on Si/SiO$_x$ at 403 K with a deposition rate of 1 Å/min and a final thickness of about 270 Å.

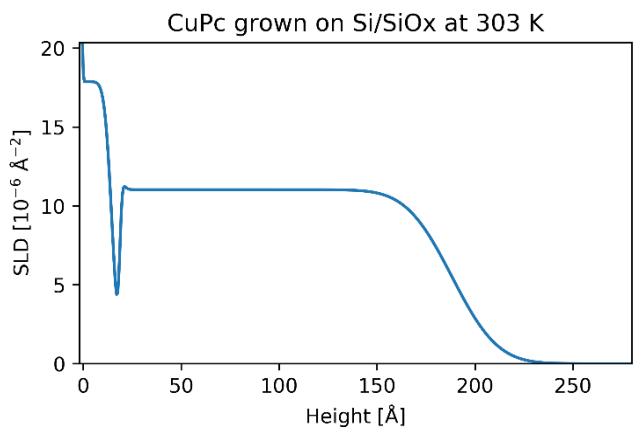

**Figure S8** SLD profile of the CuPc film grown on Si/SiO$_x$ at 303 K with a deposition rate of 0.4 Å/min and a final thickness of about 150 Å. A void layer was included between the substrate and the film to model the lower electron density at the substrate interface.

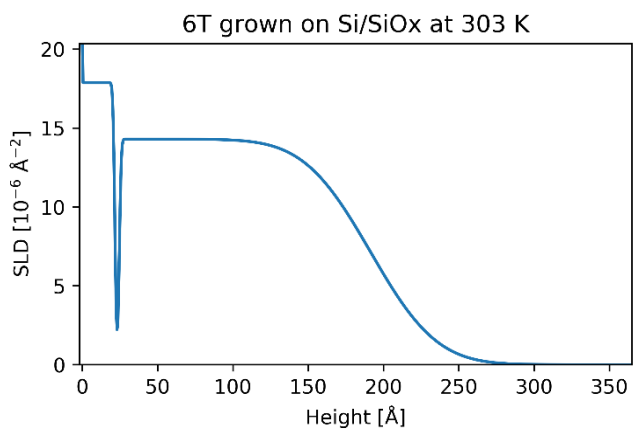

**Figure S9** SLD profile of the 6T film grown on Si/SiO$_x$ at 303 K with a deposition rate of 1.3 Å/min and a final thickness of about 165 Å. A void layer was included between the substrate and the film to model the lower electron density at the substrate interface.